**The importance of accounting for the number of co-authors and their order when assessing research performance at the individual level in the life sciences**[1]

*Giovanni Abramo*[a,b,*], *Ciriaco Andrea D'Angelo*[b], *Francesco Rosati*[b]

[a] Institute for System Analysis and Computer Science (IASI-CNR)
National Research Council of Italy

[b] Laboratory for Studies of Research and Technology Transfer
School of Engineering, Department of Management
University of Rome "Tor Vergata"

**Abstract**

Accurate measurement of research productivity should take account of both the number of co-authors of every scientific work and of the different contributions of the individuals. For researchers in the life sciences, common practice is to indicate such contributions through position in the authors list. In this work, we measure the distortion introduced to bibliometric ranking lists for scientific productivity when the number of co-authors or their position in the list is ignored. The field of observation consists of all Italian university professors working in the life sciences, with scientific production examined over the period 2004-2008. The outcomes of the study lead to a recommendation against using indicators or evaluation methods that ignore the different authors' contributions to the research results.

---

**\* Corresponding author:** Dipartimento di Ingegneria dell'Impresa, Università degli Studi di Roma "Tor Vergata", Via del Politecnico 1, 00133 Rome - ITALY, tel/fax +39 06 72597362, giovanni.abramo@uniroma2.it

# 1. Introduction

Evaluating the productive efficiency of research organizations and individual scientists is an exercise that is as important as it is delicate. The principle indicator of efficiency for almost any activity is the labor productivity, or in very simple terms, the relationship between output produced in a defined period and the hours of labor expended to produce it. As for any measurement system, that for research productivity is subject to limits and approximations, which must be duly taken into account considering the field and the intended use of the results. In particular, research activity has certain characteristics that make it notably complicated to carry out accurate and robust measurement of labor productivity. We first observe the intangible nature of the output, and also consider that such outputs are generally obtained through collaboration of various individuals, who may or may not be from the same organization or nation, and who may cooperate by contributing resources, experience and competencies that are both similar and complementary. In evaluating the scientific activity of a researcher or organization it is thus fundamental to identify the true contribution that the individual or institution has provided to the various research results in which they have from time to time participated. In the scientific fields where codification of results is primarily through publication in scientific journals, indexed in such databases as Web of Science (WoS) or Scopus, bibliometrics can be conveniently applied for large scale evaluation of productivity. In this case, the contribution of scientists and organizations to the individual publications can be recognized through the analysis of co-authorships.[2] In the life sciences, in particular, widespread practice is for the authors to indicate the various contributions to the results of the published research by the positioning of the names in the authors list.

In this work, we propose to measure the distortions encountered in the evaluation of research productivity for single individuals in Biology and Medicine when no consideration is given to the co-authors of a research work or to their order in the list.

As much as taking account of both of these factors in comparative measurement of research productivity would seem logical, and even mandatory under the theory of production, it is not at all rare that they are partially or completely ignored. In national research evaluation exercises with peer-review techniques, this is standard practice: for example in the UK Research Assessment Exercises (RAE) and in the Italian Triennial Evaluation Exercise (VTR), the peer evaluators are only called to judge the level of excellence of the products that the researchers submit, independent of true entity of the author's contribution to their accomplishment. The same is true of the national exercises that, while conducted with bibliometric techniques, examine only a share of the entire output (see the current Research Quality Evaluation exercise, VQR, in Italy). Even famous and widely used bibliometric performance indicators, such as the *h*-index (Hirsch, 2005) and the *g*-index (Egghe, 2006), totally ignore any consideration of the contributions of the individual authors to the scientific product. Little attention has been paid to advice from the inventors themselves, such as that from Hirsch (2005), who warned that "subfields with typically large collaborations (e.g., high-energy experiment) will exhibit larger *h* values", and further recommended that "in cases of large differences in the number of co-authors, it may be useful in comparing different individuals to normalize *h* by a factor that reflects the average number of co-authors".

---

[2] While there may be a human tendency to assume the opposite, it should be noted that the quality of publications is obviously a priori independent of the number of authors.

Little attention has also been paid to the specific corrections proposed, such as the simple division of the *h*-index by the average number of co-authors included in the Hirsch core (Batista et al., 2006; Egghe, 2008; Schreiber, 2009, 2010), or consideration of the actual number of co-authors and the scientists' relative position in the byline (Wan et al., 2008). In spite of the above intrinsic limits, we still see major bibliometric databases such as WoS and Scopus provide the *h*-index of every author, and it is this that scientists widely use to compare their personal performance against that of their peers, to the point that this index has now become the regulated reference threshold for access to a professorial career in Italy, both for candidates and for members of the national competition commissions (Ministerial decree 344, 4 August 2011).

In the literature, various scholars have addressed the theme of the analysis of co-authorship in evaluating scientists' research performance. Van Hooydonk (1997) pointed out that the impact of a research unit can dramatically be affected by the counting procedures. Carbone (2011) holds that "in general fractional counting is preferred because this does not increase the total weight of a single paper", and suggests that "the best way to define a fractional counting of authorship is to divide the number of citations received by each paper by the square root of the number of co-authors". As early as 1968, Zuckerman studied the patterns of name ordering in cases of multiple authorship involving Nobel laureates, and concluded that "ordering of author's names is an adaptive device which symbolizes their relative contributions to research". Based on a random selection of 5,686 chemistry papers from Current Contents volumes, Vinkler (2000) observed "only a slight preference for the alphabetical listing of authors over other rankings". In a previous work, Lukovits and Vinkler (1995) suggested that co-authors should declare their individual contributions to the research as percentages, and also introduced a simple equation for calculating individual contribution scores for coauthors of multi-authored papers. More recently Verhagen et al. (2003) proposed a Quantitative Uniform Authorship Declaration (QUAD) System that permits the reader to rapidly identify who contributed what. According to Bhandari et al. (2003) "the answer, in the tradition of scientific transparency, is for authors to decide together their individual contributions and disclose these to their readers". The author order "can reveal subtle patterns of scientific collaboration and provide insights on the nature of credit assignment among co-authors" (He et al., 2012). Trueba and Guerrero (2004) proposed a formula that assigns relative values to each co-author according to their position in the list. Laurance (2006) suggests that "the individual making the greatest intellectual contribution is the lead author, followed sequentially by those making progressively lesser contributions. In addition, the final-author slot is sometimes reserved for a lab head or project initiator, who may have made little direct contribution to the paper but deserves some vague honor nonetheless". In practice, different patterns are followed in ordering the authors list, from simple alphabetical order to sequences that signal the varying importance of the contributions from individual authors, a pattern which is particularly common in the life sciences.

There is increasing agreement among bibliometricians on the desirability of taking account of co-authorship through fractional counting, though there are still differences over the most appropriate fraction to assign to each co-author.

This work is not precisely concerned with establishing the most appropriate value to assign to contributions from co-authors in the life sciences. Rather after choosing fixed, but potentially "fine-tunable", criteria to assign different weight to the various positions in the list, the objective we set is to measure the extent of the distortion in performance ranking when the number of co-authors and their order are totally ignored. In Italy, there are no fixed guide-lines establishing the order of names in the authors list for the life sciences, even though some important academic lobbying bodies have officially pronounced themselves in favor[3]. The Italian National University Council states that the medical sciences are characterized by "scientific works that are prevailingly by multiple authors, in which the first and last authors are generally the leader of the specific research and the leader of the entire research group, and where in certain fields the second name indicates the co-leader of the specific research". In effect, wide-spread practice is that the position of first author falls to the "idea generator" and person who executes the bulk of the work, while the last position is assigned to the overall working-group leader. In the case of multiple authors from more than one institution, with similar contributions to the research, the indication of second and second-last authors also becomes significant. In general, if the position of the first author is assigned to one organization, the last name listed will be that of the group leader from the other institution, and the positions of second and second-last authors are then assigned in the opposite manner.

In the current work we will calculate the research performance of individual professors and draw up a total of six ranking lists, three for each of two types of bibliometric indicators, based on number of publications and number of citations: i) a list that takes account of both number of co-authors of each publication and their position in the list; ii) a list that does not consider position; iii) one that does not consider co-authorship in any way. The field of observation is the 2004-2008 research production by professors in Biology and Medicine from all Italian universities.

The next section of our paper illustrates the methodology and dataset used for the analyses. Section 3 presents the results concerning the correlations between the ranking lists, the analysis of shifts in position in the classifications, and a deeper examination concerning the "above-median" and top 10% of scientists. The work concludes with a summary of the results and the authors' considerations.

## 2. Methodology

### 2.1 Measuring research productivity

Research activity is a production process in which the inputs consist of human, tangible (scientific instruments, materials, etc.) and intangible (accumulated knowledge, social networks, etc.) resources, and where outputs have a complex character of both tangible nature (publications, patents, conference presentations, databases, protocols, etc.) and intangible nature (tacit knowledge, consulting activity, etc.). The new-knowledge production function therefore has a multi-input and multi-output character. The principal efficiency indicator of any production system is labor productivity. To calculate it we need to adopt a few simplifications and assumptions. In the hard sciences, including life sciences, the prevalent form of codification of research output is publication in scientific journals. As a proxy of total output in this work we consider only publications (articles, article reviews and proceeding papers) indexed in the WoS. The other forms of output which we neglect are often followed by publications that describe their content in the scientific arena, so the analysis of publications alone actually avoids a potential double counting.

---

[3] http://www.cun.it/media/100033/area6.pdf, last access Oct. 17, 2012.

When measuring labor productivity, if there are differences in the production factors available to each scientist then one should normalize by them. Unfortunately, relevant data are not available at individual level in Italy. The first assumption then is that resources available to professors within the same field of observation are the same. The second assumption is that the hours devoted to research are more or less the same for all professors. In Italy the above assumptions are acceptable because in the period of observation, core government funding was input oriented and distributed to satisfy the resource needs of each and every university in function of their size and activities. Furthermore, the hours that each professor has to devote to teaching are established by national regulations and are the same for all. As noted above, research projects frequently involve a team of researchers, which shows in co-authorship of publications. Productivity measures then need to account for the fractional contributions of scientists to their outputs. In the life sciences, the position of co-authors in the list reflects the relative contribution to the project and needs to be weighted accordingly. Furthermore, because the intensity of publications varies across fields (Abramo et al., 2008), in order to avoid distortions in productivity rankings, one must compare researchers within the same field. A prerequisite of any distortion-free research performance assessment is thus a classification of each researcher in one and only one field. In fact, in the Italian university system all professors are classified in one field. This feature of the Italian higher education system is unique in the world. In the hard sciences, there are 205 such fields (named scientific disciplinary sectors, SDSs[4]), grouped into nine disciplines (named university disciplinary areas, UDAs[5]). Since it has been demonstrated that productivity of full, associate and assistant professors is different (Abramo et al., 2011), and academic rank determines differentiation in salaries, comparisons of research performance should be differentiated by academic rank.

**2.2 Indicators**

A very gross way to calculate the average yearly labor research productivity is to simply measure the weighted fractional count of publications per researcher in the period of observation and divide it by the full-time equivalent of work in the period. A more sophisticated way to calculate productivity recognizes the fact that publications, embedding the new knowledge produced, have different values. Their value depends on their impact on scientific advancements. As proxy of impact, bibliometricians adopt the number of citations for the researchers' publications.

However, comparing researchers' performance by field and academic rank is not enough to avoid distortions in rankings. In fact citation behavior also varies across fields, and it has been shown (Abramo and D'Angelo, 2011) that it is not unlikely that researchers belonging to a particular scientific field may also publish outside that field (a typical example is statisticians, who may apply their theory to medicine, physics, social sciences, etc.). For this reason we standardize the citations for each publication[6] accumulated at June 30, 2009 with respect to the median[7] for the distribution of citations for all the Italian publications of the same year and the same subject category[8].

---

[4] The complete list is accessible on http://attiministeriali.miur.it/UserFiles/115.htm, last accessed Oct. 17, 2012.
[5] Mathematics and computer sciences; physics; chemistry; earth sciences; biology; medicine; agricultural and veterinary sciences; civil engineering; industrial and information engineering.

We consider two types of average yearly productivity measures at the individual level: a gross one based on publication counts, named weighted fractional output, WFO; and a more sophisticated one based on field-normalized citations, named weighted fractional impact, WFI. In formulae:

$$WFO = \frac{1}{t} \cdot \sum_{i=1}^{N} w_i \qquad [1]$$

Where:

$w_i$ = weight as co-author of publication *i*. Different weights are given to each co-author according to their position in the list and the character of the co-authorship (intra-mural or extra-mural). If first and last authors belong to the same university, 40% of the publication is attributed to each of them; the remaining 20% are divided among all other authors. If the first two and last two authors belong to different universities, 30% of the publication is attributed to first and last authors; 15% of the publication is attributed to second and second-last author; the remaining 10% is divided among all others[9].

N = number of publications of the researcher in the period of observation.
t = number of years of work of the researcher in the period of observation.

$$WFI = \frac{1}{t} \cdot \sum_{i=1}^{N} \frac{c_i}{Me_i} * w_i \qquad [2]$$

Where:

$c_i$ = citations received by publication *i*;
$Me_i$ = median of the distribution of citations received for all Italian cited-only publications of the same year and subject category of publication *i*;
N = same as above;
$w_i$ = same as above:
t = same as above.

Based on the above indicators, we measure a further four: two, FO and FI, eliminating the weighting that takes account of the position in the list of co-authors; and two, O and I, eliminating the fractional count that takes account of co-authors. For each indicator, we then elaborate professor ranking lists for each SDS and academic rank. To compare productivity of professors belonging to different SDSs and academic ranks, we express their productivity on a percentile scale of 0-100 (worst to best) for comparison with the performance of all Italian colleagues of the same academic rank and SDS.

---

[6] While Vinkler (2012) supports the "ratio of the sums" method, we have always preferred the "new crown indicator", even before it was corrected by the CWTS bibliometricians after criticism from Opthof and Leydesdorff (2010) concerning the statistical normalization of the "old" indicator.

[7] We standardize citations by the median, because as frequently observed in literature (Lundberg, 2007), standardization of citations with respect to median value rather than to the average is justified by the fact that distribution of citations is highly skewed in almost all disciplines. However, there is not yet agreement among bibliometrician on the most efficient scaling factor.

[8] The subject category of a publication corresponds to that of the journal where it is published. For publications in multidisciplinary journals the scaling factor is calculated as a weighted average of the standardized values for each subject category.

[9] The weighting values for both this indicator and the WFI indicator below were assigned based on the results of interviews with top Italian professors in the life sciences. The values could be changed to suit different practices in other national contexts.

**2.3 Dataset**

Data on research staff of each university and their SDS classification are extracted from the database on Italian university personnel, maintained by the Ministry for Universities and Research[10]. The bibliometric dataset used to measure productivity is extracted from the Italian Observatory of Public Research (ORP)[11], a database developed and maintained by the authors and derived under license from the Thomson Reuters WoS. Beginning from the raw data of the WoS, and applying a complex algorithm for reconciliation of the author's affiliation and disambiguation of the true identity of the authors, each publication is attributed to the university scientist or scientists that produced it (D'Angelo et al., 2010).

We use two datasets for our analysis, both built by beginning with all professors that meet the following two conditions: i) they belong to the SDSs in the Biology and Medicine UDAs (Appendix A), where bibliometric techniques provide a robust calculation of productivity[12] and where the number of professors per academic rank is equal to or greater than 10; and ii) they held their position for at least three years during the period 2004-2008. The datasets then differentiate for a third condition. For the analysis of WFO, the condition is that the professors had at least one publication during the period, while for the analysis of WFI it is that their overall publications achieved at least one citation. In fact, for the purposes of our project it would not make sense to consider professors with nil output or citations, given that this means nil productivity, independent of the choice of indicator. Overall, the first dataset includes 13,658 professors belonging to 63 universities, and the second counts 13,392 professors belonging to 62 universities. The SDSs analyzed are 19 in Biology and 43 in Medicine (Table 1).

*Table 1: Datasets for the analysis*

| UDA | SDSs | Publications* | Citations | WFO | | WFI | |
|---|---|---|---|---|---|---|---|
| | | | | Professors | Universities | Professors | Universities |
| Biology | 19 | 27,600 | 218,105 | 4,718 | 62 | 4,652 | 61 |
| Medicine | 43 | 50,331 | 407,311 | 8,940 | 53 | 8,740 | 53 |
| Total | 62 | 70,740** | 563,201** | 13,658 | 63 | 13,392 | 62 |

*Number of publications authored by at least one academic professor of the UDA*
** *The value differs from the sum of the two previous lines as a result of multiple counts related to publications coauthored by both Biology and Medicine professors.*

---

[10] http://cercauniversita.cineca.it/php5/docenti/cerca.php. Last accessed on Oct. 17, 2012.
[11] www.orp.researchvalue.it. Last accessed on Oct. 17, 2012.
[12] To ensure the representativity of publications as proxy of the research output, the field of observation was limited to those SDSs where at least 50% of researchers produced at least one WoS-indexed publication in the period 2004-2008.

## 3. Results

In this section we will present the results from the comparisons between the different productivity ranking lists. Comparisons are between the ranking lists derived from the same type of productivity indicator: those based on simple publication count (WFO, FO, O) and those based on standardized citations (WFI, FI, I). The ranking lists are prepared for each SDS and academic rank. We begin with the correlation analyses between the ranking lists; continue with the analyses of shifts in position when changing from ranking under one indicator to rankings under another, and conclude with a deeper analysis of the shifts in position for the above-median and top 10% of scientists.

### 3.1 Correlation analysis

The correlation analyses between the ranking lists for each of the three impact-based productivity indicators show significant and strong correlation for all three comparisons (Table 2). The highest overall correlation is between indicators FI and I (0.956 for general correlation, 0.949 for Biology and 0.960 for Medicine), followed by the correlation between WFI and FI (0.922 for general correlation, 0.931 for Biology and 0.917 for Medicine) and WFI and I (0.872 for general correlation, 0.884 for Biology and 0.869 for Medicine).

*Table 2: Correlation analyses between ranking lists for productivity indicators based on impact, per UDA and SDS*

|  |  | WFI – I | WFI – FI | FI – I |
|---|---|---|---|---|
| Biology | Observations | 4,652 | 4,652 | 4,652 |
|  | General correlation | 0.879 | 0.932 | 0.949 |
|  | Max correlation | 0.950 (BIO/15) | 0.977 (BIO/05) | 0.979 (BIO/15) |
|  | Min correlation | 0.764 (BIO/18) | 0.886 (BIO/18) | 0.884 (BIO/18) |
| Medicine | Observations | 8,740 | 8,740 | 8,740 |
|  | General correlation | 0.869 | 0.917 | 0.960 |
|  | Max correlation | 0.914 (MED/14) | 0.953 (MED/39) | 0.977 (MED/24) |
|  | Min correlation | 0.721 (MED/22) | 0.762 (MED/22) | 0.923 (MED/32) |
| Total | Observations | 13,392 | 13,392 | 13,392 |
|  | General correlation | 0.872 | 0.922 | 0.956 |

The ranking lists for productivity indicators based on publication count again show significant and strong correlations (Table 3). The highest average correlation is for indicators FO and O (0.937 for general correlation, 0.920 for Biology and 0.946 for Medicine), followed by the correlation between WFO and FO (0.923 for general correlation, 0.925 for Biology and 0.922 for Medicine) and WFO and O (0.866 for general correlation, 0.856 for Biology and 0.871 for Medicine).
We observe that in both cases, the weakest correlation is between the "complete" indicator and the one that ignores both the co-authors and their position in the list.

*Table 3: Correlation analyses between ranking lists for productivity indicators based on publication count, per UDA and SSD*

|  |  | WFO – O | WFO – FO | FO – O |
|---|---|---|---|---|
| Biology | Observations | 4,718 | 4,718 | 4,718 |
|  | General correlation | 0.856 | 0.925 | 0.920 |
|  | Max correlation | 0.928 (BIO/07) | 0.976 (BIO/07) | 0.972 (BIO/15) |
|  | Min correlation | 0.705 (BIO/08) | 0.873 (BIO/12) | 0.745 (BIO/08) |
| Medicine | Observations | 8,940 | 8,940 | 8,940 |
|  | General correlation | 0.871 | 0.922 | 0.946 |
|  | Max correlation | 0.943 (MED/16) | 0.962 (MED/16) | 0.972 (MED/24) |
|  | Min correlation | 0.670 (MED/46) | 0.765 (MED/46) | 0.905 (MED/37) |
| Total | Observations | 13,658 | 13,658 | 13,658 |
|  | General correlation | 0.866 | 0.923 | 0.937 |

## 3.2 Analysis of changes in position under ranking lists for different indicators of impact

In this section we compare the ranking lists for each of the three impact indicators, observing the changes in position of the scientists in each UDA. In Biology (Figure 1), 54.8% of changes in position between the FI and I ranking lists are distributed within the interval [0;5].

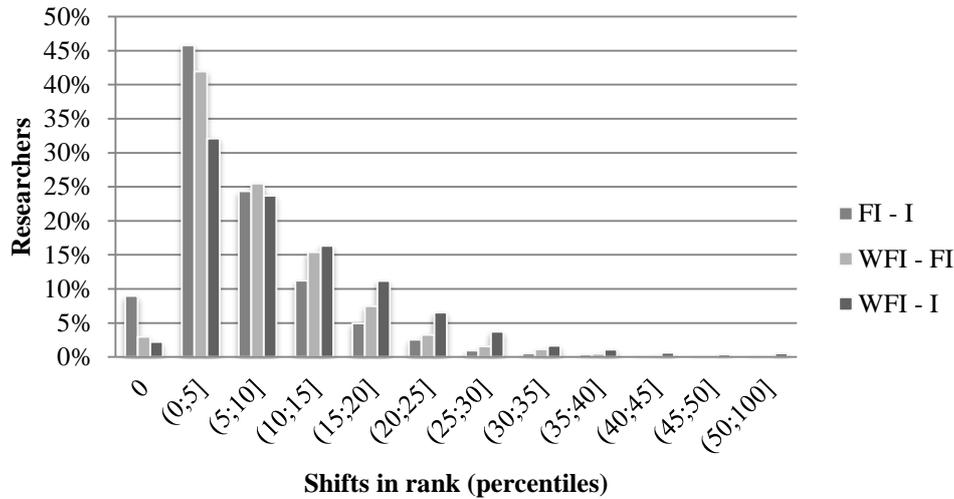

*Figure 1: Distributions of changes in position for Biology professors under FI and I ranking lists, under WFI and FI lists, and WFI-I lists*

For the same interval, the changes in position between WFI and FI lists descend to 44.9%, and for WFI-I to 34.3%. We also observe that the FI-I changes in list position feature a very high peak corresponding to the lowest shifts and a quite short right tail corresponding to increasing values of shifts. The highest shifts in position appear with greatest frequency in the comparison between WFI and I ranking lists.

In Medicine (Figure 2), 59.4% of the shifts in position between FI and I fall within the interval [0;5]; the shifts in the same interval for the WFI-FI comparison drop to 42.5%, and for WFI-I to 33.9%. We also observe that the distribution of shifts between FI and I features a still higher peak than Biology, in correspondence with the lowest values of shift, with a quite short right tail corresponding to increasing values of shift. Again as in Biology, the larger shifts in position occur with greatest frequency in the comparison between WFI and I ranking lists.

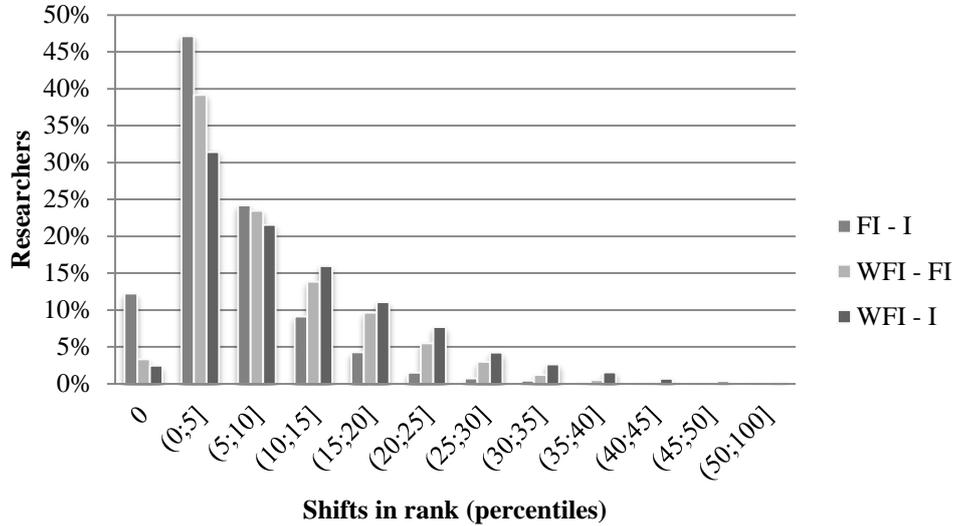

*Figure 2: Distributions of changes in position for Medicine professors under FI and I ranking lists, under WFI and FI lists, and WFI-I lists*

We next compare only the ranking lists for WFI and I, in both Biology and Medicine (Table 4), measuring the percentage of scientists who classify in a different quartile (1 to 4 scale, best to worst quartile). We observe that in Biology, 35.0% of scientists would change quartile: 4.5% would drop from first under the WFI ranking list to second quartile under the I list; 6.7% would drop from second to third quartile, and 6.4% from third to fourth. Shifts of two quartiles are very rare, with the most frequent (1.0%) being from the fourth to second quartile.

*Table 4: Percentage of scientists who change quartile (1 to 4, best to worst) when changing from WFI to I ranking list*

|     |   | Biology |      |      |      |   | Medicine |      |      |      |
|-----|---|---------|------|------|------|---|----------|------|------|------|
|     |   | I       |      |      |      |   | I        |      |      |      |
|     |   | 1       | 2    | 3    | 4    |   | 1        | 2    | 3    | 4    |
| WFI | 1 | 19.8    | 5.4  | 0.2  | 0.0  | 1 | 19.7     | 5.7  | 0.4  | 0.0  |
|     | 2 | 4.5     | 13.2 | 6.7  | 0.4  | 2 | 4.4      | 12.6 | 7.3  | 0.6  |
|     | 3 | 0.7     | 4.9  | 13.3 | 6.4  | 3 | 1.1      | 5.1  | 12.0 | 7.3  |
|     | 4 | 0.2     | 1.0  | 4.6  | 18.6 | 4 | 0.0      | 1.1  | 5.1  | 17.7 |

Medicine registers still more relevant shifts, with 38.0% of scientists shifting a quartile: 5.7% would drop from first to second, 7.3% from second to third, and 7.3&  from third to fourth quartile. A two quartile shift occurs for a maximum of 1.1% of scientists.

We next conduct a finer analysis of the shifts in position between the percentile ranking lists for all the impact indicators, in each SDS of the two disciplines. Table 5 presents the descriptive statistics for the SDSs that register the maximum values. In Biology, the highest average value of percentile shift (10.6) occurs in the comparison between the ranking lists for WFI and I, and the SDS with the highest average shift value (14.3) is BIO/02 (Systematic botany). The highest single shifts are in BIO/18 (Genetics), with a shift of 81.5 between WFI and I, and 53.8 between WFI and FI, meaning that high performers under one ranking list would be low in the other list. In Medicine, the shifts in position between WFI and I are more accentuated than in Biology. The highest overall average percentile shift (11.2) occurs in the comparison between these lists, and the highest average shift for an individual SDS (17.2, for MED/22 - Vascular surgery) also occurs between WFI and I. Still comparing the WFI and I lists, the maximum shift (57.5) occurs in MED/30 (Eye diseases), while in the comparison between WFI and FI lists the maximum shift (56.7) is in MED/06 (Medical oncology).

*Table 5: Descriptive statistics for distributions of percentile differences of professors' productivity rankings, per impact indicator*

|  |  | WFI – I | WFI – FI | FI – I |
|---|---|---|---|---|
| Biology | Avg. shift in rank | 10.6 | 7.8 | 6.5 |
|  | Max avg. shift in rank | 14.3 (BIO/02) | 10.1 (BIO/19) | 9.2 (BIO/18) |
|  | Max shift in rank | 81.5 (BIO/18) | 53.8 (BIO/18) | 67.7 (BIO/18) |
|  | Min Stand. Dev. of shifts in rank | 6.1 (BIO/15) | 4.2 (BIO/05) | 4.4 (BIO15) |
|  | Max Stand. Dev. of shifts in rank | 14.5 (BIO/18) | 9.8 (BIO/18) | 11.0 (BIO/08) |
| Medicine | Avg shift in rank | 11.2 | 8.8 | 5.6 |
|  | Max avg shift in rank | 17.2 (MED/22) | 15.7 (MED/22) | 8.1 (MED/32) |
|  | Max shift in rank | 57.5 (MED/30) | 56.7 (MED/06) | 65.0 (MED/36) |
|  | Min Stand. Dev. of shifts in rank | 8.0 (MED/14) | 6.2 (MED/17) | 4.4 (MED/40) |
|  | Max Stand. Dev. of shifts in rank | 14.1 (MED/22) | 13.3 (MED/22) | 8.9 (MED/32) |

## 3.2 Analysis of changes in position under ranking lists for different indicators of output

In this section we repeat the same analyses as above, but now for the indicators of output. In Biology (Figure 3), 42.3% of the shifts in position between the ranking lists for FO and O fall within interval [0;5], the shifts in the same interval for the WFO to FO comparison drop to 46.5%, and for WFO-O to 32.7%. We also observe that the distribution of shifts between FO and O features a peak corresponding to the lowest values of shift and a quite short right tail corresponding to increasing values of shift. The highest shifts in position occur with greater frequency in the comparison between the WFO and O ranking lists.

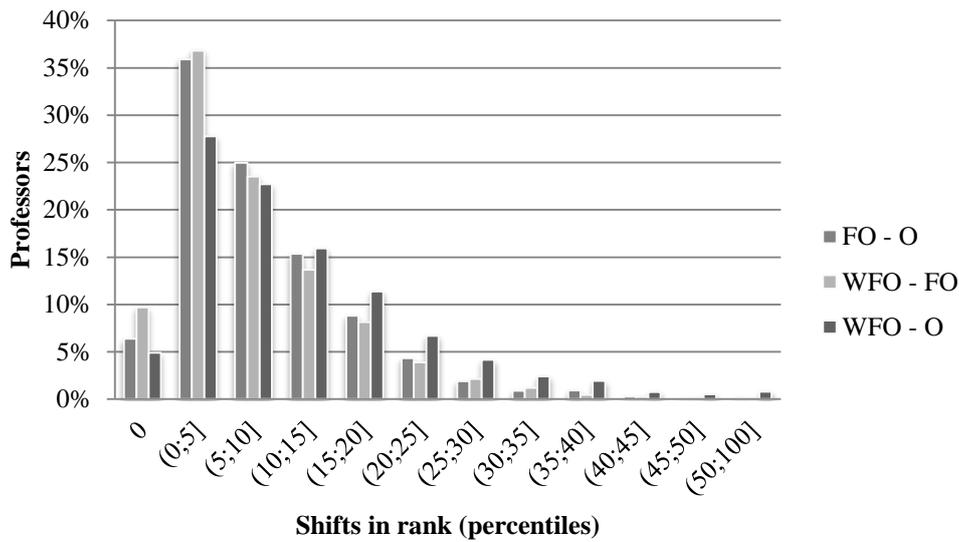

*Figure 3: Distributions of changes in position for Biology professors under FO and O, WFO-FO and WFO-O ranking lists*

In Medicine (Figure 4), 48.0% of the shifts in position between FO and O ranking lists fall within interval [0;5]; the shifts in the same interval for the WFI-FI comparison drop to 44.8%, and for WFO-O to 33.7%. We also observe that the distribution of shifts between FO and O features a peak in correspondence with the lowest values of shift, with a quite short right tail in correspondence with increasing values of shift. The greatest shifts in position occur with more frequency in the comparison between the WFO and O ranking lists.

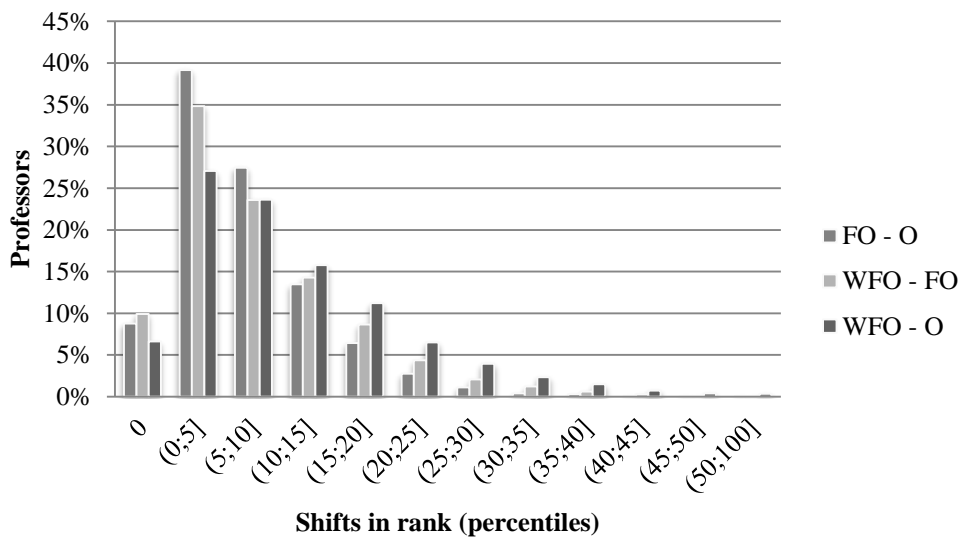

*Figure 4: Distributions of changes in position for Medicine professors under FO and O, WFO-FO and WFO-O ranking lists*

We next compare the ranking lists for WFO and O in Biology and Medicine, measuring the percentage of scientists that classify in a different quartile (Table 6). We see that in Biology, 37.8% of scientists would change quartile: 4.4% would drop from first quartile under the WFI ranking list to second under I; 5.8% would drop from second to third, and 4.9% from third to fourth quartile. Shifts of two quartiles are very rare, with the most frequent (1.6%) being between fourth and second.

In Medicine, 35.6% of scientists would change quartile: 4.3% drop from first to second, 5.6% from second to third, and 4.9% from third to fourth quartile. A two quartile shift occurs for a maximum of 1.4% of scientists, with the jump from fourth to second quartile.

*Table 6: Percentage of scientists who change quartile when changing from WFO to O ranking list*

|     |     | Biology |      |      |      |     | Medicine |      |      |      |
|-----|-----|---------|------|------|------|-----|----------|------|------|------|
|     |     |         | O    |      |      |     |          | O    |      |      |
|     |     | 1       | 2    | 3    | 4    |     | 1        | 2    | 3    | 4    |
| WFO | 1   | 20.3    | 4.4  | 0.4  | 0.1  | 1   | 20.6     | 4.3  | 0.3  | 0.0  |
|     | 2   | 5.0     | 13.2 | 5.8  | 0.5  | 2   | 4.9      | 13.4 | 5.6  | 0.4  |
|     | 3   | 1.0     | 7.1  | 12.0 | 4.9  | 3   | 0.7      | 6.2  | 13.0 | 4.9  |
|     | 4   | 0.2     | 1.6  | 6.9  | 16.7 | 4   | 0.0      | 1.4  | 6.8  | 17.2 |

We next conduct a finer analysis of the shifts in position between the percentile ranking lists for all the indicators of output, in each SDS of the two disciplines. Table 7 presents the descriptive statistics for the SDSs that register the maximum values. In Biology, the highest average value of shift in percentile (11.6) occurs in the comparison between ranking lists for WFO and O, and the SDS with the highest average value of shift (17.7%) is BIO/18 (Anthropology). Still comparing the WFO and O lists, the maximum shift (73.8) occurs in BIO/18 (Genetics), while in the comparison between WFO and FO the maximum shift (60.0) occurs in BIO/12 (Clinical biochemistry and molecular biology). In Medicine, the highest overall average shift in percentile (11.0) is seen in the comparison of lists for WFO and O, and the SDS with the highest average value of shift (19.7) is MED/46 (Medical laboratory technique). In the comparison between ranking lists for WFO and O, the maximum shift (68.0) occurs in MED/22 (Vascular surgery), while in the WFO to FO comparison the maximum shift (55.6) is in MED/23 (Cardiac surgery).

*Table 7: Descriptive statistics for distributions of percentile differences of professors' productivity rankings, per output indicators*

|          |                                | WFO – O        | WFO – FO       | FO – O         |
|----------|--------------------------------|----------------|----------------|----------------|
| Biology  | Avg. shift in rank             | 11.6           | 8.0            | 8.6            |
|          | Max avg. shift in rank         | 17.7 (BIO/08)  | 10.4 (BIO/04)  | 15.7 (BIO/08)  |
|          | Max shift in rank              | 73.8 (BIO/18)  | 60.0 (BIO/12)  | 72.3 (BIO/18)  |
|          | Min Stand. Dev. of shifts in rank | 7.6 (BIO/07) | 4.4 (BIO/07)   | 5.6 (BIO/15)   |
|          | Max Stand. Dev. of shifts in rank | 13.5 (BIO/08)| 10.7 (BIO/12)  | 13.5 (BIO/08)  |
| Medicine | Avg. shift in rank             | 11.0           | 8.3            | 7.1            |
|          | Max avg. shift in rank         | 19.7 (MED/46)  | 16.5 (MED/46)  | 9.4 (MED/37)   |
|          | Max shift in rank              | 68.0 (MED/22)  | 55.6 (MED/23)  | 64.3 (MED/04)  |
|          | Min Stand. Dev. of shifts in rank | 6.8 (MED/16)| 5.6 (MED/16)   | 5.0 (MED/35)   |
|          | Max Stand. Dev. of shifts in rank | 14.6 (MED/46)| 12.6 (MED/46) | 9.7 (MED/37)   |

## 3.4 Analysis of changes in position for above-median and top 10% of scientists

In this section we focus on the changes in position for two particularly important subgroups of scientists: those that place above the median or in the top 10% of the ranking list. We see which percentage of scientists in each subgroup would change position in a ranking list when switching from one indicator to another. For the indicators of impact (Table 8), switching from the ranking list for WFI to the one for simple I, 31.8% of the top 10% of scientists in biology and 31.3% of those in Medicine would no longer be "top", while 14.6% of those above median in Biology and 16.4% of those in Medicine would drop below. The like percentages are reduced in the comparison based on the other indicators.

For the indicators of output (Table 9), the highest percentages of scientists who would no longer be top 10% or above-median are again registered in the switch from the ranking list for WFO to that for O. A full 31.0% of top 10% researchers in Biology and 26.7% of those in Medicine would no longer be "top", while 13.6% of the above-median in Biology and 12.8% of those in Medicine would drop below median.

*Table 8: Percentages of above-median and top 10% scientists who do not remain such with change in the impact indicator*

|  | Biology | | Medicine | |
|---|---|---|---|---|
| from – to | No longer top 10% scientists (%) | No longer above median (%) | No longer top 10% scientists (%) | No longer above median (%) |
| WFI – I | 31.8 | 14.6 | 31.3 | 16.4 |
| WFI – FI | 19.8 | 10.7 | 21.3 | 13.1 |
| FI – I | 24.9 | 8.7 | 20.6 | 7.8 |

*Table 9: Percentages of above-median and top 10% scientists who do not remain such with change in the output indicator*

|  | Biology | | Medicine | |
|---|---|---|---|---|
| from – to | No longer top 10% scientists (%) | No longer above median (%) | No longer top 10% scientists (%) | No longer above median (%) |
| WFO – O | 31.0 | 13.6 | 26.7 | 12.8 |
| WFO – FO | 17.8 | 11.7 | 19.9 | 11.2 |
| FO – O | 27.3 | 9.0 | 20.6 | 7.3 |

## 3.5 Conclusions

There is a rapidly increasing tendency to evaluate the research activity of individual scientists, research groups, and entire organizations, as a support instrument for improving efficiency in national research systems. Evaluation at the individual level informs recruitment, incentive systems and selective resource allocation. Bibliometrists are thus called on to refine techniques and indicators that can render the evaluation tools every more accurate and robust. The different contributions of the specific authors in realizing scientific advancement through co-authorship must certainly be taken account in individual evaluation. In the life sciences, there are widespread and consolidated practices to signal the different contributions of the individual authors to the research results. Many bibliometric indicators and national evaluation exercises ignore this important specificity of the life sciences, failing to consider the order of the co-authors and even their number. In this work we have indicated an order of magnitude for the distortion in performance ranking that occurs under such circumstances.

The benchmark indicators we used to measure labor productivity are the less refined weighted fractional output, WFO, and the more sophisticated weighted fractional impact, WFI. Beginning from each of the main indicators, we measured another four: two, FO and FI, eliminate the weighting that takes account of position in the co-authors list; a further two, O and I, eliminate the fractional count that provides for the number of co-authors. Comparison to the ranking list for each indictor, for each field of research and academic rank, permitted us to reveal the shift from the respective benchmarks.

The extent of distortion is considerable, even for the indicators that retain consideration of co-authors but do not take account of their different contributions: in the case of the measures based on impact indicators (comparison WFI-FI), the average shift in percentile rank is 7.8 for Biology and 8.8 for Medicine, with peaks of maximum shift in rank of up to 53.8 (in BIO/01) for Biology and 56.7 (in MED/06), for Medicine.

The distortions are still more evident when the indicators do not take account of the number of co-authors (comparison WFI-I): the average shift in percentile rank is 10.6 for Biology and 11.2 for Medicine, with peaks of maximum rank shift equal to 81.5 in BIO/18 and 57.5 in MED/30. Not taking account of the number of co-authors would result in a full 31.8% of the top 10% scientists in Biology and 31.3% of those in Medicine no longer being recognized as such, while 14.6% of above-median researchers in Biology and 16.4% of those in Medicine would then be classified as below the median. Similar values are seen in the comparison based on indicators of output.

In a context where collaboration in research is ever more the norm (98.2% of Italian university articles in Biology and Medicine are in co-authorship and 93.6% show more than two authors), it is evident that using bibliometric indicators that ignore the contribution of each individual author leads to the introduction of distortions that can: a) notably reduce the accuracy and reliability of the measure; b) undermine the effectiveness of the entire evaluation process; and, according to the uses made, c) compromise the results at the micro- and macro-economic level.

The evidence from this study should inspire caution concerning the use of widely-diffused indicators in the scientific world (such as the *h*-index and others) that do not take account, among other considerations, of the number of co-authors. The same can be said for peer-review methodologies applied to the evaluation of individuals and organizations that, in the analysis of the quality of a share of scientific production (such as in the UK RAE and others), then ignore the order and number of co-authors in the products evaluated.

# Appendix A

| UDA | SDS_code | SDS_title |
|---|---|---|
| | MED/01 | Medical Statistics |
| | MED/02 | History of Medicine* |
| | MED/03 | Medical Genetics |
| | MED/04 | General Pathology |
| | MED/05 | Clinical Pathology |
| | MED/06 | Medical Oncology |
| | MED/07 | Microbiology and Clinical Microbiology |
| | MED/08 | Pathological Anatomy |
| | MED/09 | Internal Medicine |
| | MED/10 | Respiratory Diseases |
| | MED/11 | Cardiovascular Diseases |
| | MED/12 | Gastroenterology |
| | MED/13 | Endocrinology |
| | MED/14 | Nephrology |
| | MED/15 | Blood Diseases |
| | MED/16 | Rheumatology |
| | MED/17 | Infectious Diseases |
| | MED/18 | General Surgery |
| | MED/19 | Plastic Surgery |
| | MED/20 | Pediatric and Infant Surgery |
| | MED/21 | Thoracic Surgery |
| | MED/22 | Vascular Surgery |
| Medicine | MED/23 | Cardiac Surgery |
| | MED/24 | Urology |
| | MED/25 | Psychiatry |
| | MED/26 | Neurology |
| | MED/27 | Neurosurgery |
| | MED/28 | Odonto-Stomalogical Diseases |
| | MED/29 | Maxillofacial Surgery |
| | MED/30 | Eye Diseases |
| | MED/31 | Otorhinolaryngology |
| | MED/32 | Audiology |
| | MED/33 | Locomotory Diseases |
| | MED/34 | Physical and Rehabilitation Medicine** |
| | MED/35 | Skin and Venereal Diseases |
| | MED/36 | Diagnostic Imaging and Radiotherapy |
| | MED/37 | Neuroradiology |
| | MED/38 | General and Specialized Pediatrics |
| | MED/39 | Child Neuropsychiatry |
| | MED/40 | Gynecology and Obstetrics |
| | MED/41 | Anesthesiology |
| | MED/42 | General and Applied Hygiene |
| | MED/43 | Legal Medicine* |
| | MED/44 | Occupational Medicine |
| | MED/45 | General, Clinical and Pediatric Nursing** |
| | MED/46 | Laboratory Medicine Techniques |
| | MED/47 | Nursing and Midwifery* |
| | MED/48 | Neuropsychiatric and Rehabilitation Nursing** |
| | MED/49 | Applied Dietary Sciences** |
| | MED/50 | Applied Medical Sciences |

| UDA | SDS_code | SDS_title |
|---|---|---|
| | BIO/01 | General Botany |
| | BIO/02 | Systematic Botany |
| | BIO/03 | Environmental and Applied Botany |
| | BIO/04 | Vegetal Physiology |
| | BIO/05 | Zoology |
| | BIO/06 | Comparative Anatomy and Cytology |
| | BIO/07 | Ecology |
| | BIO/08 | Anthropology |
| Biology | BIO/09 | Physiology |
| | BIO/10 | Biochemistry |
| | BIO/11 | Molecular Biology |
| | BIO/12 | Clinical biochemistry and molecular biology |
| | BIO/13 | Applied Biology |
| | BIO/14 | Pharmacology |
| | BIO/15 | Pharmaceutical Biology |
| | BIO/16 | Human Anatomy |
| | BIO/17 | Histology |
| | BIO/18 | Genetics |
| | BIO/19 | General Microbiology |

*\* SDSs excluded from the database because bibliometric techniques are not sufficiently robust to calculate productivity*
*\*\* SDSs excluded because they have less than 10 professors for each academic rank*